# Drop-Wise and Film-Wise Water Condensation Processes Occurring on Metallic Micro-Scaled Surfaces

Anton Starostin[a], Viktor Valtsifer[a], Zahava Barkay[b], Irina Legchenkova[c], Viktor Danchuk[d], Edward Bormashenko[c]

[a]*Institute of Technical Chemistry, UB RAS, Academician Korolev St., 3, 614013 Perm, Russian Federation*

[b]*Wolfson Applied Materials Research Center, Tel Aviv University, Ramat-Aviv 69978, Israel*

[c]*Engineering Faculty, Chemical Engineering, Biotechnology and Materials Department, Ariel University, Ariel 40700, Israel*

[d]*Exact Sciences Faculty, Physics Department, Ariel University, Ariel 40700, Israel*

**Abstract**

Water condensation was studied on silanized (superhydrophobic) and fluorinated (superoleophobic) micro-rough aluminum surfaces of the same topography. Condensation on superhydrophobic surfaces occurred via film-wise mechanism, whereas on superoleophobic surfaces it was drop-wise. The difference in the pathways of condensation was attributed to the various energy barriers separating the Cassie and Wenzel wetting states on the investigated surfaces. The higher barriers inherent for superoleophobic surfaces promoted the drop-wise condensation. Triple-stage kinetics of growth of droplets condensed on superoleophobic surfaces is reported and discussed.

## 1. Introduction

Nano- and micro-scaled surfaces are of much industrial interest in a view of their numerous applications as water-repellent (self-cleaning) [1-5], superoleophobic [6-7] and ice-phobic surfaces [8-10]. Such structured surfaces also provide drag reduction and the pressure drop in pipes [11-13]. Nano- and micro-scaled metallic meshes have been successfully used for oil/water separation [14-17]. These prepared with common metals such as steel [9, 17-19], copper, brass [9, 20-22] and aluminum [23-25] are of a practical industrial importance. Our paper is focused on water condensation on micro-scaled aluminum surfaces, which is of particular interest for preventing corrosion [24]. Aluminum is inherently a hydrophilic material, however, manufacturing of appropriate surface reliefs may convert aluminum surfaces into hydrophobic and even superhydrophobic ones [25].

Water condensation on nano- and micro-scaled surfaces has been studied intensively in the last decade [26-32]. Water condensation could be generally classified as drop-wise [26-32] and film-wise [33]. Drop-wise condensation on a hydrophobic surface has been known to enhance heat transfer by approximately an order of magnitude compared to film condensation on a hydrophilic one [34-35]. So, the issue of manufacturing metallic surfaces keeping their hydrophobic properties in a course of condensation is thus of primary technological importance. Note, that a few natural and artificial nanostructured superhydrophobic surfaces retain their superhydrophobic characteristics during water condensation [32]. It should be emphasized that physical processes taking place under condensation, including wetting transitions, nucleation and kinetics of droplets' growth, are complicated and not well understood [28]. The role of the hierarchy of spatial scales on the dropwise condensation on the superhydrophobic surfaces was discussed [32]. The impact of local energy barriers on the process of condensation was elucidated [30]. In our investigation, we focus on the possibility of fine tuning of surface properties of aluminum micro-rough surfaces enabling switching from film-wise to drop-wise pathway of water condensation and also on the kinetics of growth of droplets.

**2. Experimental: materials and methods.**

*2.1. First stage in preparation of superhydrophobic and superoleophobic surfaces.*

Micro-scaled metallic surfaces were prepared as follows. Aluminum plates were thoroughly cleaned with ethanol and acetone. The cleansed aluminum plates were immersed for 5 min in a 5% water solution of hydrochloric acid (HCl was supplied by Alfa Aesar). Etching of the aluminum plates by the HCl gave rise to the micro-rough relief depicted in **Figure 1**.

*2.1.1. Preparing of superhydrophobic surfaces.*

Superhydrophobic (abbreviated SH) metallic surfaces were prepared as follows. The samples were dried during 10 min at 100 °C after the etching stage (Section 2.1). The dried micro-rough surface of aluminum plates was hydrophobized with 3-5% solution of Polymethylhydrosiloxane (PMHS, supplied by Sigma Aldrich, MW=222.5) in organic solvent Hexanes (supplied by Alfa Aesar). The mass ratio of PMHS to the silica specimen equaled 1:10. After removal of the solvent, the specimens were dried at 100 °C for 10 minutes, and subsequently thermally treated at 200 °C for 3 hours. The surface after modification acquired SH properties demonstrating the water apparent contact angle $\theta^* = 152 \pm 1^0$ (see **Figure 2A**) and low contact angle hysteresis studied in detail in Ref. 25.

*2.1.2. Preparing of superoleophobic surfaces.*

Superoleophobic metallic surfaces (abbreviated below SO) were prepared as follows. The samples were dried during 10 min at 100°C after the etching stage (Section 2.1). Dried micro-rough aluminum plates were immersed for 30 min in a solution of perfluorononanoic acid, 97% $C_9HF_{17}O_2$ (supplied by Alfa Aesar). Then the hydrophobized plates were dried during 30 min at 80 °C. After modification, the surface acquired superoleophobic properties, with a high value of the apparent contact angle, established as $\theta^* = 155 \pm 1^0$ for water (see **Figure 2B**) and oils established as $\theta^* = 153 \pm 1^0$ for canola oil and $\theta^* = 155 \pm 1^0$ for castor oil. Low contact angle hysteresis and high stability of the Cassie wetting states typical for the superoleophobic surfaces were registered [25, 35]. The apparent contact angles were measured by a Ramee-Hart Advanced Goniometer Model 500-F1 under ambient conditions.

*2.2. ESEM study of the condensation observed on superhydrophobic and superoleophobic surfaces.*

The structures were studied in the Quanta 200FEG ESEM (Environmental Scanning Electron Microscope) while imaging the sample surface using the gaseous SE (secondary electron) detector. In-situ imaging of water condensation was performed within the ESEM at wet-mode using water vapor from a built-in distilled water reservoir (specific resistivity $\rho = 40 k\Omega \times cm$). At the beginning of the experiment, the Peltier stage, which held the sample, was stabilized at 2°C while the vapor pressure in the chamber was held at 5.3 Torr. The relative humidity in the chamber was increased by elevating the pressure above dew point (to 6-7 Torr) for in-situ imaging of water condensation.

*2.3. Study of the stability of the Cassie wetting states on the superhydrophobic and superoleophobic surfaces.*

Study of the stability of the Cassie wetting states on the SH and SO surfaces was carried out under the protocol described in detail in Refs. 19, 25, 36. Droplets of water/ethanol mixtures with various concentrations of ethanol were placed on the surfaces and the apparent contact angle was taken. When the concentration of ethanol in the droplet was increased the apparent contact angle consequently decreased. This procedure enabled fixation of the onset of the Cassie-Wenzel transition, and the estimation of the stability of the Cassie wetting state [35, 37-41].

**3. Results and discussion.**

*3.1. Study of the condensation on micro-scaled superhydrophobic and superoleophobic surfaces.*

ESEM study shows the very different pathways of condensation processes on the micro-scaled chemically modified SH and SO surfaces. The film-wise condensation was observed on the SH surfaces as shown in **Figure 3 (a1)**. Starting from the 25$^{th}$ second of the condensation (see Section 2.2) the sections of water films covering the surfaces have been registered. At the long scale of *ca* 10 minutes the entire surface of the SH surfaces was covered with a water film, as depicted in **Figure 3 (a2)**. In contrast, the process of condensation on the SO surface proceeded via the drop-wise mechanism. **Figure 3 (b1)** depicts appearance of micro-sized spherical water drops condensed on the SO surface. The condensed droplets grow and coalesce; the kinetics of growth will be discussed in detail in Section 3.2. At the long time scale of *ca* 10 minutes, the droplets grew from the micro- to the millimeter size and have been detached (see **Figure 3 (b2**) and Refs. 26-27, 42-43).

The main question is: what is the physical reasoning leading to the different mechanisms of condensation on the SH and SO surfaces, demonstrating the same topography of **Figure 1**? The answer to this question is crucial for design of ice-phobic and optimal heat transfer surfaces [8-10, 29, 44]. Note, that the values of water apparent contact angles on both of the surfaces were very close, as discussed in Sections 2.1.1-2.1.2. It is reasonable to relate the observed difference in the processes of condensation to various stability of the Cassie wetting state [37-41, 45], inherent for the studied SH and SO surfaces. The stability of the Cassie wetting regime was estimated by the establishment of the critical surface tension of a droplet $\gamma_c$, corresponding to the onset of the Cassie-Wenzel transition (see Section 2.3 and Refs. 19, 25).

The curves $\theta^*(\gamma)$ (where $\theta^*$ is the apparent contact angle as measured on SH and SO surfaces and $\gamma$ is the surface tension of a droplet at the liquid-air interface) are presented in **Figure 4**. It is seen that the SO curve demonstrates a more stable Cassie wetting; i.e. the critical surface tension $\gamma_c$, corresponding to the onset of the SH (~50 mJ/m$^2$) surfaces higher than that for SO ones (~40 mJ/m$^2$). Hence, it seems plausible the relate the drop-wise condensation observed on SO surfaces to the higher stability of its Cassie wetting regime, arising from fluorination process, involved in its manufacturing (see Section 2.1.2). We conclude that the fluorination of aluminum surfaces supplies to them the higher stability of Cassie wetting in comparison to their silanization. Our study supports the idea that local energy barriers, separating the Cassie and Wenzel wetting states play a crucial role in the constituting the regime of condensation, suggested in Refs. [30, 45].

Note that the difference in the stability of the Cassie wetting for SO and SH surfaces is minor, as recognized from the curves, shown in **Figure 4**. Hence, even a tenuous difference in the value of a potential barrier separating the Cassie and Wenzel wetting states may be crucial for constituting the regime of condensation. The pure qualitative characterization of the stability of Cassie wetting according to $\theta^*(\gamma)$ curves should be emphasized; and further quantitative investigations in this field are required.

*3.2. Kinetics of the droplets' growth.*

Nucleation and growth of droplets on micro- and nano-rough surfaces was recently explored both theoretically [46-47] and experimentally [48], showing that the topography and the specific surface energy of the substrates strongly affect nucleation and kinetics of droplets' growth. Triple-stage kinetics of the droplet growth under dropwise condensation, observed on SO, is illustrated with **Figure 5**. Let us start from the eventual stage of the process corresponding to coalescence of droplets followed by their detachment. The detachment stage ("jumping") was addressed in detail in Refs. 26-27, 42-43. Self-propelled detachment of condensate drops is driven by the surface energy released upon drop coalescence [26-37, 42-43].

It is seen from **Figure 5** that at the time of $t_0 = 52.5s$ the kinetics of growth was switched (the radius of a growing droplet denoted $r_0$ at that time was approximately $r_0 = 5.0 \mu m$). We approximated the dependence $r(t)$ (where $r$ is the radius of a droplet) with power functions $r(t) \sim t^\alpha$, as suggested in Refs. 49, 50.

At the initial stage, the growth is well approximated by Eq. 1 (see the double logarithmic dependence $\frac{r(t)}{r_0}$ vs $\frac{t}{t_0}$, presented in **Figure 5b**):

$$0 < t < t_0, \qquad r(t) \sim t^{0.2} \quad , \tag{1}$$

The "experimental" value of exponent, namely $\alpha \cong 0.20$ coincides satisfactorily with that reported in Ref. 49 (namely $\alpha \cong 0.23$); however, it was much smaller than values of exponents, established for high and low-pinned droplets in Ref. 50 ($\alpha \cong 0.78$ and $\alpha \cong 0.46$, correspondingly). The established value of the exponent at the initial stage of condensation is close to that reported in Ref. 52 for the growth of asymmetric droplets. Note also the appearance of asymmetric droplets in our experiments, recognized in **Figure 3(1b)**. It is noteworthy that the precise value of the exponent corresponding to the initial stage of a droplet growth remains debatable [49-52]. The decrease of the growth exponent in a course of a droplet growth at constant water vapor pressure and temperature was reported in Ref. 53.

The initial stage of a growth was followed by the stage, starting at $t_0 \cong 52.5 \pm 0.5 s$ satisfactorily approximated by Eq. 2 (see **Figure 5b**):

$$t_0 < t < t_1, \; r(t) \sim t^{0.92} \qquad (2)$$

The established experimentally value of exponent ($\alpha \cong 0.92$) is close to the values predicted for the growth of droplets accompanied with their coalescence and also for the environmental situation, when the velocity of the atmosphere surrounding a droplet is very small ($0.75 < \alpha < 1$), as shown in Refs. 49-51, 54. It is reasonable to suggest that in the ESEM experiments the growing droplet "swallows" not only individual molecules of water, but also growing water clusters, dispersed in the atmosphere of the ESEM chamber. This "pseudo-coalescence growth" results in the high values of the exponent α, extracted from the experimental data, related to the stage of the process antecedent to the droplets' detachment.

Now, consider the numerical value of the radius of a growing droplet $r_0 = 5.0 \mu m$, at which the switch from "low" values of the exponent ($\alpha \cong 0.20$) to "large" ($\alpha \cong 0.92$) ones takes place. Note, that this switch occurs, when the radius of a droplet is close to the mean free path of water molecules λ, calculated as follows:

$$\lambda \cong \frac{k_B T}{\sqrt{2}\pi d^2 P}, \qquad (3)$$

where $k_B$ = 1.38×10$^{-23}$ J/K is the Boltzmann constant, $T$ = 276.4±0.1 K and $P$ = 840±1 Pa (6.3 Torr) are the temperature and the pressure in the condensation camera corresponding to the dew point, and $d = 2.65 \overset{0}{A}$ is the kinetic diameter of a molecule of water [55]. Simplest calculation yields the rough estimation: $\lambda \cong 14.5 \pm 0.1 \mu m$. Note, that the values of $r_0$ and $\lambda$ are of the same order of magnitude.

We suggest that this proximity is not accidental. Indeed, when the interrelation $\lambda \leq r_0$ takes place, the slowing of the droplet growth is expected, due to the lower concentration of water molecules at the bottom part of the droplet, as shown in **Figure 6**. The so-called "depletion layer" is formed close to the substrate due to the different factors, including water adsorption by the substrate, and nucleation, stimulated by a rough surface [46, 54]. The concentration of water molecules in the depletion layer is smaller than that in the volume of ESEM chamber, as shown in **Figure 6**.

In a course of the droplet growth its radius becomes close to the mean free path of water molecules $\lambda$, and the influence of the depletion layer on the kinetics of a droplet growth, will decline (the specific surface of a droplet "shadowed" by the depletion layer will decrease in a course of growth); thus, promoting the increase in the growth exponent $\alpha$. Of course, the quantitative considerations resulting in the accurate estimation of the switch point of the growth exponent are desirable.

**Conclusions**

Environmental scanning electron microscopy of water condensation on silanized (superhydrophobic) and fluorinated (superoleophobic) micro-rough aluminum surfaces is reported. We observed film-wise condensation on SH surfaces and pronounced drop-wise condensation on the SO surfaces, possessing the similar topography. The stability of the Cassie wetting regime was studied on the both kinds of surfaces, within the experimental establishment of the critical value of the surface tension of a liquid giving rise to the Cassie-Wetting transition [35-41]. These measurements supplied the comparative semi-qualitative information about the stability of the Cassie wetting on the reported surfaces. The stability of the Cassie wetting on the SO surfaces, quite expectably, turned out to be higher. Higher barriers inherent for SO surfaces promoted the drop-wise condensation. We conclude that energy barriers, separating the Cassie and Wenzel wetting states play a crucial role in the constituting the regime of condensation, as suggested in Refs. 30, 56.

The triple-stage kinetics of droplets, which grew on SO surfaces, is reported. The time dependence $r(t)$ (where $r$ is the radius of a growing droplet) was approximated with power functions $r(t) \sim t^{\alpha}$ [49-54]. At the initial stage of the growth the value of exponent, ($\alpha \cong 0.20$) coincides satisfactorily with that reported in Ref. 49 and it is close to that reported in Ref. 52. At the intermediate stage of the growth the exponent switched to $\alpha \cong 0.92$. The switch in the value of the growth exponent is at least partially due to the formation of the depletion layer of water molecules adsorbed by the substrate in the vicinity of a droplet [46, 54]. We conjectured that the abrupt change in the growth exponent took place when the radius of a droplet became close to the mean free path of water molecules in the ESEM chamber.

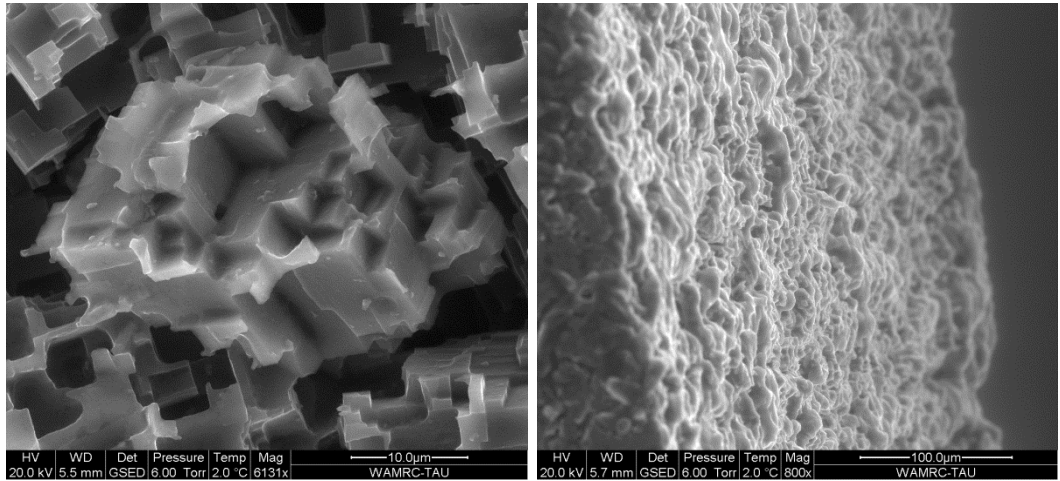

**Figure 1**. ESEM images of the micro-rough aluminum plate are depicted. **A**. Scale bar is 10µm. **B**. Scale bar is 100 µm.

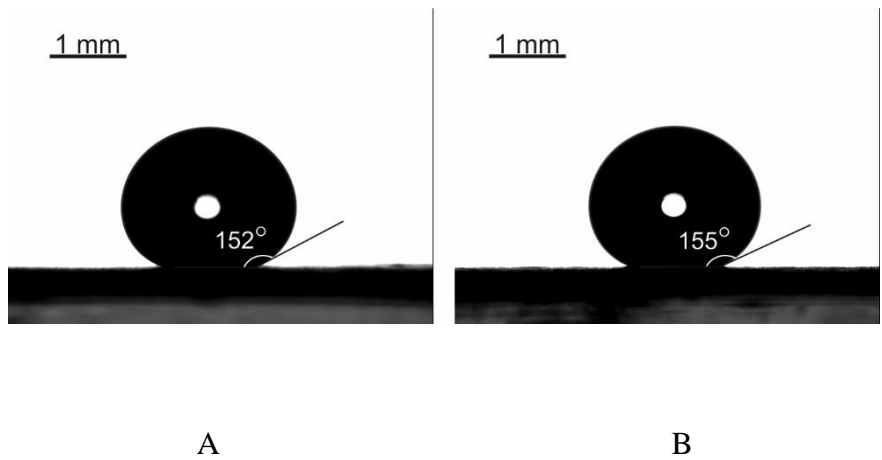

A B

**Figure 2**. Images of 8 µl water droplets placed on a superhydrophobic surface (A) and superoleophobic metallic aliminium (B) are depicted.

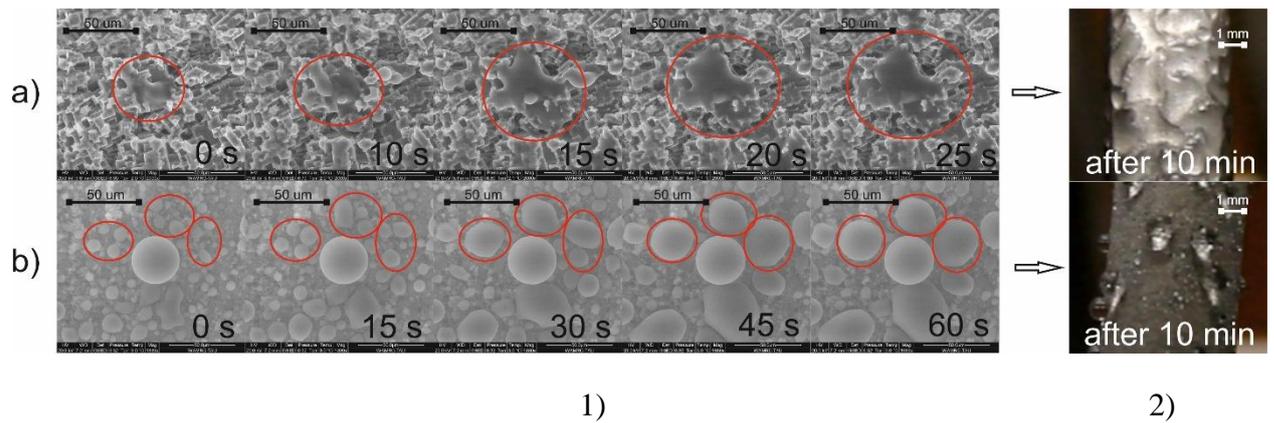

**Figure 3**. 1) ESEM images illustrating condensation of water droplets on superhydrophobic (a) and superoleophobic aluminum surfaces (b) are presented. Film-wise (a) and drop-wise (b) condensation processes are clearly recognized.

2) Eventual wetting scenarios arising from film-wise and drop-wise condensation processes observed on superhydrophobic (a) and superoleophobic (b) aluminum surfaces.

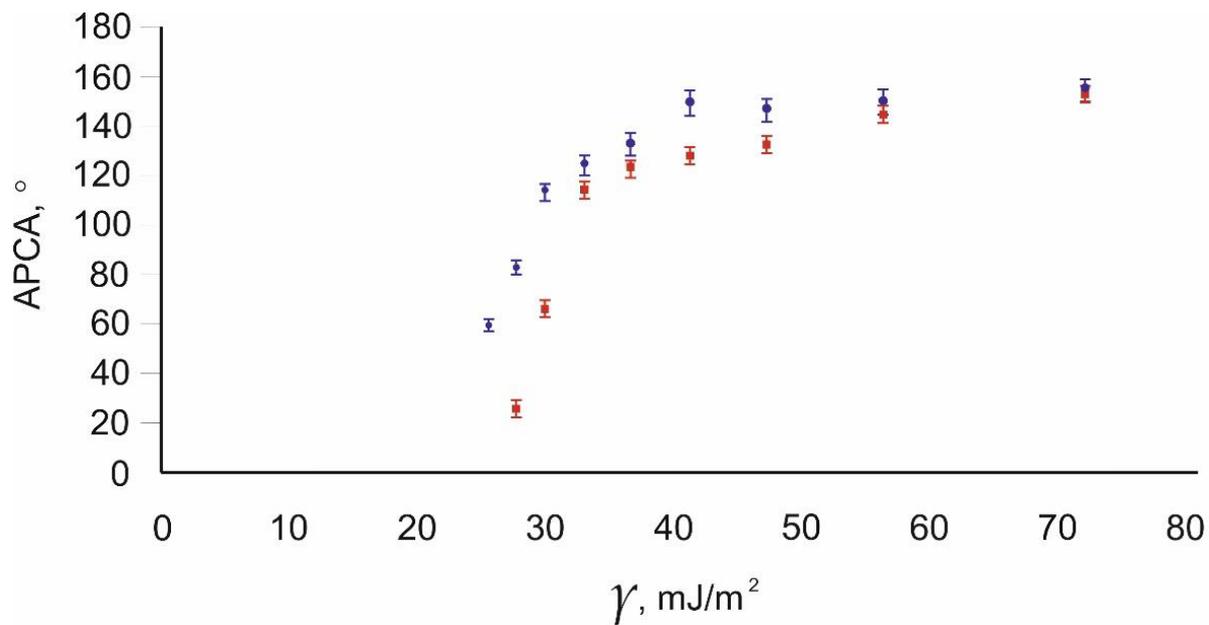

**Figure 4**. The dependencies of the apparent contact angle (APCA) *vs.* the surface tension of water/ethanol solutions are depicted. Red squares represent measurements performed with the superhydrophobic surface; blue circles correspond to the superoleophobic surface.

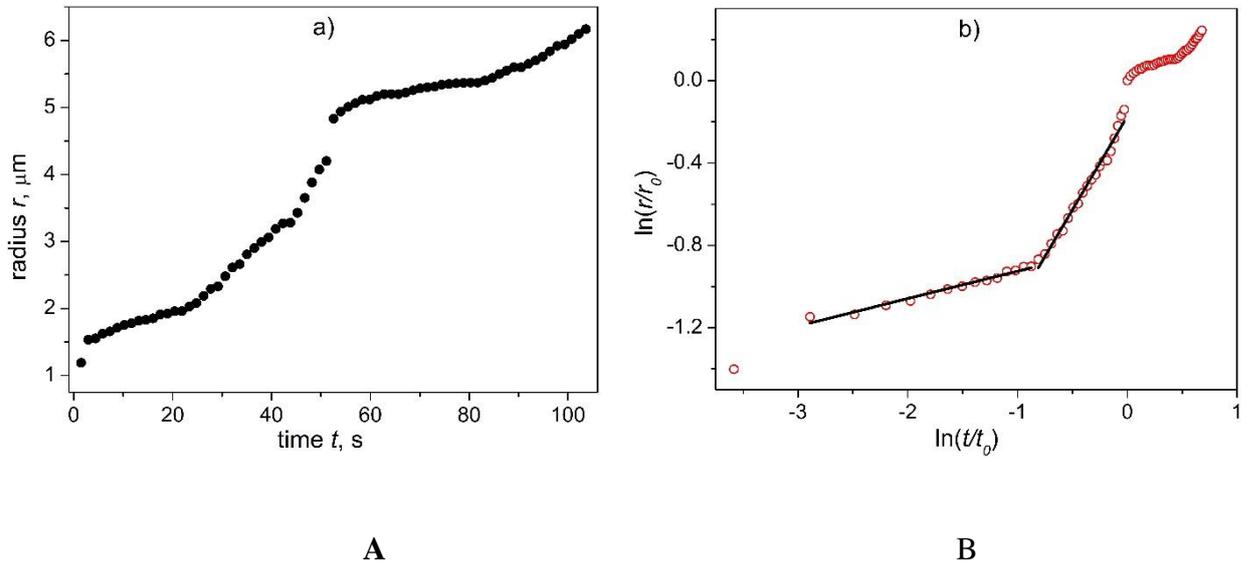

**Figure 5**. Kinetics of the droplet growth is shown. **A**. The time dependence of the droplet radius is depicted. **B**. The same dependence for dimensionless parameters $\frac{r(t)}{r_0}; r_0 = 5.0 \mu m$ and $\frac{t}{t_0}; t_0 = 52.5 s$ is presented in the double logarithmical coordinates.

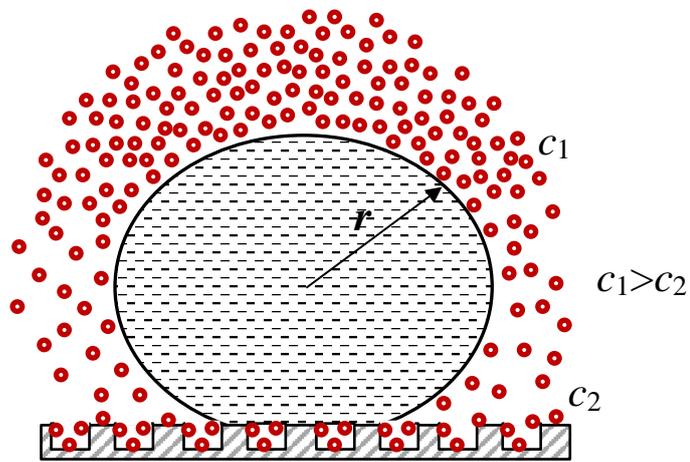

**Figure 6**. Formation of the "depletion layer" close to the micro-rough surface is shown schematically. The raduis of a condensed droplet $r$ is smaller than the mean free path of water molecules (depicted with red circles). The concentration of water molecules in the depletion layer $c_2$ is smaller than that in the volume of the condensation chamber $c_1$.